\documentclass[twocolumn,amssymb,amsmath, superscriptaddress]{revtex4}
\usepackage{graphicx}
\usepackage{amssymb}

\def\0{\varnothing}

\def\ctbp{{\it Center for Theoretical Biological Physics, \\University of California at San Diego, La Jolla, CA 92093}}

\begin{document}
\title{Correlated Phenotypic Transitions to Competence in Bacterial Colonies}
\author{Inbal Hecht}
\affiliation{\ctbp}
\author{Eshel Ben-Jacob}
\affiliation{\ctbp} \affiliation{\it School of Physics and
Astronomy, Tel-Aviv University, Israel}
\author{Herbert Levine}
\affiliation{\ctbp}

\begin{abstract}
Genetic competence is a phenotypic state of a bacterial cell in
which it is capable of importing DNA, presumably to hasten its
exploration of alternate genes in its quest for survival under
stress. Recently, it was proposed that this transition is
uncorrelated among different cells in the colony. Motivated by
several discovered signaling mechanisms which create colony-level
responses, we present a model for the influence of quorum-sensing
signals on a colony of \emph{B. Subtilis} cells during the
transition to genetic competence. Coupling to the external signal
creates an effective inhibitory mechanism, which results in
anti-correlation between the cycles of adjacent cells. We show that
this scenario is consistent with the specific experimental
measurement, which fails to detect some underlying collective
signaling mechanisms. Rather, we suggest other parameters that
should be used to verify the role of a quorum-sensing signal. We
also study the conditions under which phenotypic spatial patterns
may emerge.
\end{abstract} \maketitle

Genetic competence is a cellular differentiation process, in which
bacterial cells (such as those of {\em Bacillus Subtilis})
synthesize a specific system of proteins for the binding and uptake
of DNA~\cite{review}. Internalizing exogenous DNA leads to genetic
transformations which presumably increases survival
probability~\cite{cohen}. Under certain environmental conditions, a
population of genetically identical cells (grown in homogeneous and
identical environments) will at a given time be composed of two
distinct subpopulations, a competent minority versus a non-competent
majority, each with distinguished features. This has often been
attributed to bistability~\cite{dubnau_bi,stochastic,pos_feedback}
in the genetic regulatory network responsible for the competent
state and indeed the necessary positive feedback has been found in
the dynamics of one of the key regulatory molecules
(comK)~\cite{auto_stim,elowitz}.

More recently, Elowitz and co-workers~\cite{elowitz} have proposed
that the competence transition is reversible and that a competent
cell will return to its vegetative state after a characteristic
time. This means that one should think of the network as being
excitable~\cite{excitable}, with stochasticity (perhaps due to small
molecular numbers) occasionally driving quiescent cells into the
excited competent state. This result is very interesting, as it
represents one of the first examples where the inherent
stochasticity of genetic networks~\cite{noise,paulsson,alon,vano} is
being put to functional use. One consequence of this new picture is
that the competent subpopulation is not fixed but rather is
continually fluctuating. Surprisingly, they have also claimed that
this transition is made on the single-cell level, and is
uncorrelated among different cells. This is supported by the
measurement of the competence excursion times of neighboring cells.
But in this work we show that this may not be the right parameter to
measure, if quorum sensing is to be studied. The purpose of this
paper is to present a putative quorum-sensing mechanism that induces
short-ranged spatial anti-correlations between cells, and to suggest
the appropriate way to experimentally detect such a mechanism. This
effect can be mediated by extracellular signaling molecules that are
needed for the transition. In this scenario, a cell becoming
competent will bias its neighbors against making the same choice.
Yet, excursion times remain uncorrelated, as obtained
experimentally. Our work is motivated by the many recently
discovered instances in which signaling between bacterial
cells~\cite{bassler} are used to create colony-level collective
responses to external environmental
challenges~\cite{trends,interface}.  For instance, it is clear that
the sporulation response in bacillus is spatially
coordinated~\cite{minsky}. At the end, we will discuss experimental
tests of our predictions.

We start by considering the molecular interactions underlying the
competence circuit. The onset of competence is stimulated by cell
crowding and nutrient limitation.  Detection of cell crowding is
achieved by cell-cell signaling; various molecules are secreted and
sensed by the cells, to monitor population density (quorum sensing)
as well as other environmental conditions. In B. Subtilis, CSF
(competence and sporulation factor) is a diffusible peptide, derived
from the PhrC pheromone, which is secreted by the cells during
growth. As its extracellular concentration increases due to cell
crowding, CSF is transported back into the cell
\cite{signaling,quorum} where it influences a cascade of several
genes, leading eventually to either sporulation or competence,
depending on its concentration. For low concentrations (but above
the transport threshold) it (indirectly) enhances the
phosphorylation of ComA, and consequently the expression of ComS.
ComS competes with ComK for degradation by the MecA-ClpC-ClpP
complex, thereby causing an indirect activation of ComK by ComS.
ComK activates the transcription of several genes that encode DNA
transport proteins and is considered as the indicator for the
competent state. ComK activates its own expression and indirectly
inhibits the expression of ComS. This double positive feedback loop
can, as already mentioned, lead to a bistable response, i.e. two
possible states with different gene expression. The negative
feedback loop governs the escape from the competent state, as ComK
degradation increases when ComS production is inhibited.
\begin{figure}
\centerline{\includegraphics[scale=0.5]{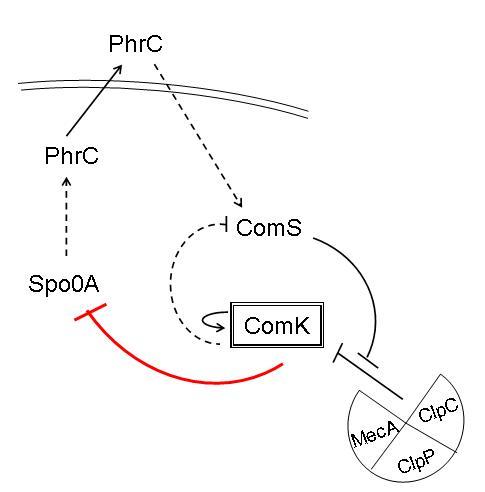}}
\caption{Schematic representation of the competence circuit in B.
subtilis. The autoregulatory positive feedback loop of ComK and the
indirect repression of ComS are presented in the right-hand side.
ComK and ComS compete for degradation by the complex MecA-ClpC-ClpP.
The quorum sensing mechanism is presented in the left-hand side.
Phrc is secreted by the cells and diffuses through the cell
membrane. As its intracellular concentration increases, it diffuses
back into the cell, initiating the competence pathway by regulating
the ComS production. ComK inhibits the production of Spo0A, which
results in the inhibition of the cell crowding signal Phrc. Solid
lines represent direct interaction, dashed lines represent indirect
interaction.}\label{model}
\end{figure}

In ref. \cite{elowitz,elowitz2} a model was proposed for the part of the
circuit downstream from the external signal transducer ComA.
Specifically, they proposed
\begin{eqnarray}
\frac{\partial K}{\partial t} & = & a_K + \frac{b_K K^n}{K_0^n +K^n} - \frac{K}{1+K+S} \nonumber \\
\frac{\partial S}{\partial t} & = &  \frac{b_S K_1^p }{K_1^p +K^p} - \frac{K}{1+K+S} +\eta (t)
\end{eqnarray}
where $K$ and $S$ represent the concentrations of the ComK and ComS
proteins.  $a_K$ (+ $b_K$) represent the minimal (fully activated)
rates of ComK production, and $K_0$  is the concentration of ComK
required for 50\% activation. The cooperativities of ComK
auto-activation and ComS repression are parameterized by the Hill
coefficients n=2 and p=5. Similarly, the expression of ComS has
maximum rate $b_S$  and is half-maximal when $K=K_1$ . Degradation
by the MecA complex affects both ComK and ComS; the nonlinear
degradation term corresponds to such a competitive mechanism. Random
fluctuations in ComS expression are represented by a noise term
$\eta(t)$. Degradation rates of K and S are normalized to unity and
are assumed to be identical. The parameters can be chosen to give
rise to an excitable system where there exists a stable fixed point
plus two other intersections of the K-S nullclines,  a saddle point
and an unstable fixed point. This case is presented in Fig. 2(a),
where we also show a typical trajectory for which the noise has
caused a large excursion, i.e. has excited the competent state.
Eventually the cell returns to the vegetative stable state, due to
the repression of ComS by ComK; when the value of S decreases below
a threshold level, K is more rapidly degraded and can no longer
auto-activate its production. One can obtain similar results with more biologically realistic noise forms~\cite{stochastic, elowitz2,schultz}.
\begin{figure}
\centerline{\includegraphics[scale=0.27]{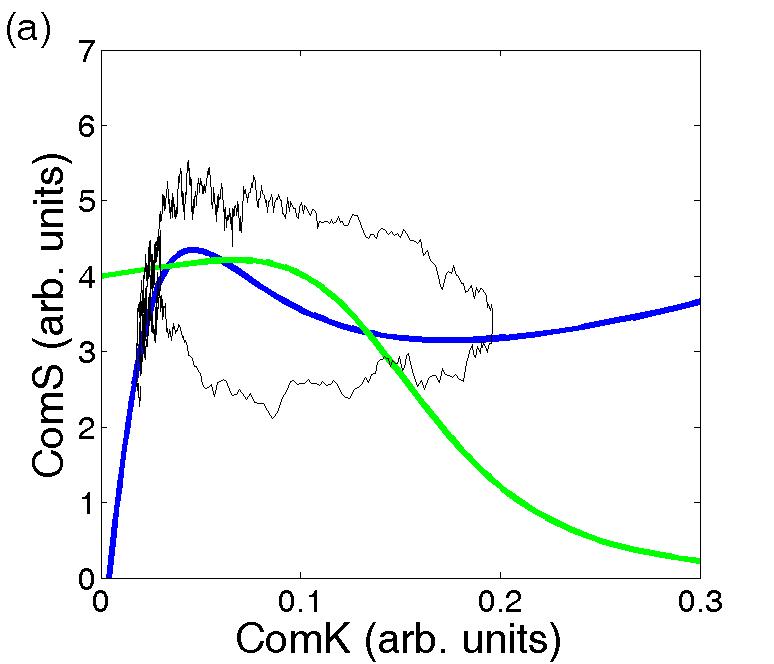}}\centerline{\includegraphics[scale=0.27]{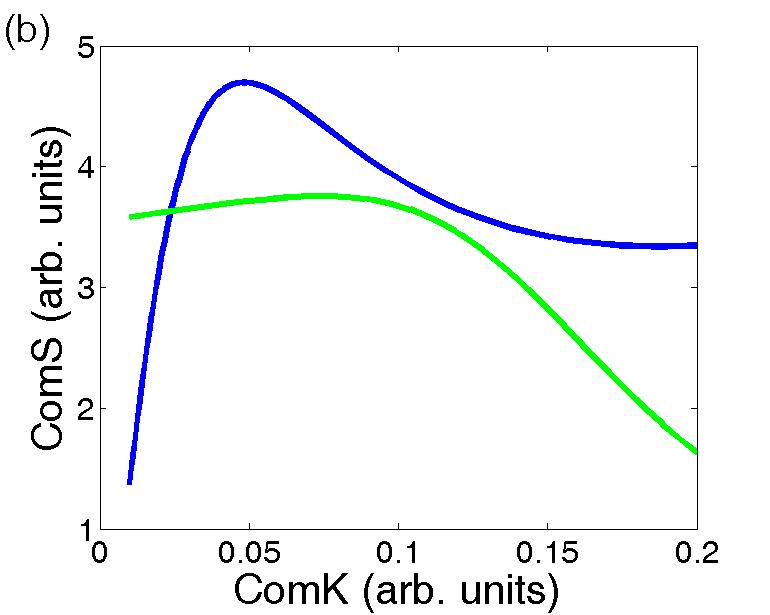}}
\caption{Phase planes diagram for the ComK-ComS system (Eqs.
(1)-(2)). Null clines are presented for ComK and ComS, and the
system fixed points. (a) The excitable system, obtained from Eqs.
(1)-(2) with the parameters: $a$=0.004, $b_{K}$=0.08, $b_{S}$=0.8,
$k_{0}$=0.2, $k_{1}$=0.222. The noise RMS amplitude is 3.5 and
$\Delta t$ =0.01. The fixed points are (from left to right) a stable
point, a saddle point and an unstable point. A characteristic
trajectory is presented for a single cell, staring from the
vegetative state (low ComK). (b) The monostable system, for Eq. (3)
and $Q$=0.9, $b_{s}$=0.6, $c_{s}$=0.2, $\delta_{Q}$=1, $D_{Q}$=1
(other parameters as in (a)). The only fixed point refers to the
vegetative state.}\label{phase-plane}
\end{figure}

In this model, each cell was taken to be independent of the other
cells. This assumption cannot be completely correct, however, since
the external enabling signal acting through ComA comes from the rest
of the colony. Obviously, the functional utility of having a
competent subpopulation would clearly be enhanced if this
subpopulation were spatially dispersed. In order to consider the
possible spatial effect of quorum sensing, we introduce an extension
of the model, and take into account a putative biological mechanism
(see Fig. 1), in which the basal production level of ComS depends on
a quorum sensing pheromone Q. The motivation for this particular model is a
microarray analysis  \cite{microarray} which has shown that the gene
Spo0A, which is upstream of the PhrC secretion mechanism, is down
regulated when ComK is high. Thus, the secretion of CSF, which is a
regulator for competence onset, is effectively inhibited by ComK.
Specifically, we assume that
\begin{equation}
\frac{\partial S}{\partial t}  =   c_S Q(x,t) +\frac{b_S K_1^p }{K_1^p +K^p} - \frac{K}{1+K+S} +\eta (t)
\end{equation}
Now $K$ and $S$ are functions of space as well as time, and $Q$
depends on the cells' state in the following manner:
\begin{equation}
\frac{\partial Q}{\partial t} = \frac{1}{2} \tanh{ (A(K_{max}-K)+1)} +D \nabla^2 Q - \delta_Q Q
\end{equation}
The production of Q is maximal as $K \rightarrow 0$  , namely normal
growth, and it decreases as the cell becomes competent. $K_{max}$ is
chosen to be the threshold value of $K$, used to define the
competent state. $Q$ diffuses with diffusion coefficient $D$  , and
is degraded with a constant rate $\delta_Q$. The parameters are
chosen to describe a bistable system for $Q=1$, an excitable system
for  $Q $ close to but less than one and a monostable one if $Q
<<1$. In fact, for the system described in Fig. 2(a), $Q$ level as
low as $0.9$ is sufficient to drive the system to the vegetative
monostable regime, as can be seen in Fig. 2(b). This is a key point,
as very small changes in the external signal (e.g. the level of CSF)
may drive the cell into a completely different regime. Therefore,
even a 2-fold change in the production of Spo0A, and as a result in
the secretion of CSF, may have a substantial effect on a possible
competence transition.

Our goal in this model is a semi-quantitative evaluation of
correlations in the competence status of nearby cells. We will study
this question in one spatial dimension (with periodic boundary
conditions), with a cell at every one of the 5000 grid points. In
each explicit time step, the values of $K$ and $S$ are updated for
each cell and then $Q$ is updated in the entire space.
\begin{figure}
\centerline{\includegraphics[scale=0.4]{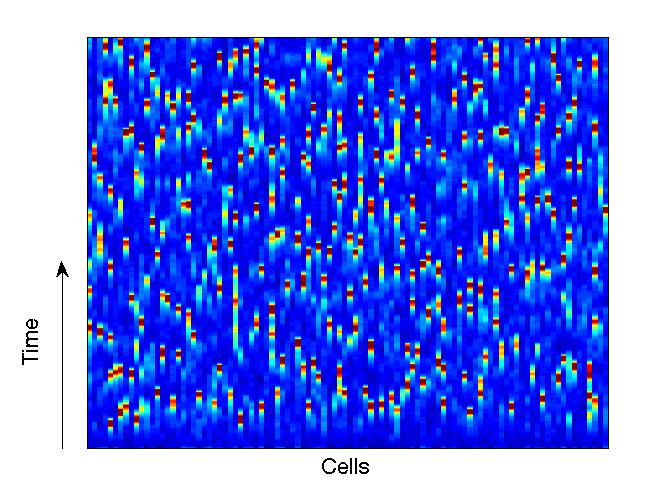}} \vspace{-.5cm}
\caption{Time development of a colony of cells. Color code
represents the ComK level, from low ComK level (vegetative state,
black) to high ComK level (competent state, white). The system
parameters are as in Fig. 1(b). All cells are initially vegetative,
and competence events occur with an average waiting time of 45
hours. The average excursion time in the competent state is about 5
hours. Adjacent cells rarely become competent simultaneously, due to
effective mutual inhibition.}\label{space-time}
\end{figure}
A typical simulation run of our model starting from all vegetative
cells is shown in Fig. 3. There are clear anti-corelations between
cells, due to the effective inhibition created by a reduction of $Q$
emission by the competent cells. To get a more quantitative measure,
we measured the spatial correlation function
\begin{equation}
C_d (i) \equiv \frac {\left<(K_i(t)- \bar{K}_i) ( K_{i+d} (t) - \bar{K} _{i+d}) \right>}{\sigma _i \sigma_{i+d}}
\end{equation}
where we average over time, $\sigma$ is the standard deviation, and
the subscript refers to a specific cell $i$. We then average over
all cells $i$ to obtain an estimate $\bar{C} _d$. In Fig. 4a  we
present the function $C_1$  , i.e. the correlation between adjacent
cells, versus the diffusion coefficient  of the quorum sensing
pheromone $Q$.  For a small diffusivity, the inhibiting effect is
negligible as the diffusion length is small compared to the
inter-cell distance.  As  $D$ is increased, the mutual inhibition is
increased and the correlation is negative, as expected. The maximal
effect on nearest neighbors is when the characteristic length
$\lambda=\sqrt{D \tau}$, given by the diffusivity and the
degradation time $\tau=\delta_Q^{-1}$ is comparable to the
inter-cellular distance. For the given parameters, the maximal
effect on adjacent cells, i.e. $\lambda$=1, is obtained for $D=1$.
For an inter-cellular distance of $1\mu m$, the highest mutual
inhibition on nearest neighbors will be obtained for $D=1/\tau$
$(\mu m^{2}/sec)$.
\begin{figure}
\centerline{\includegraphics[scale=0.27]{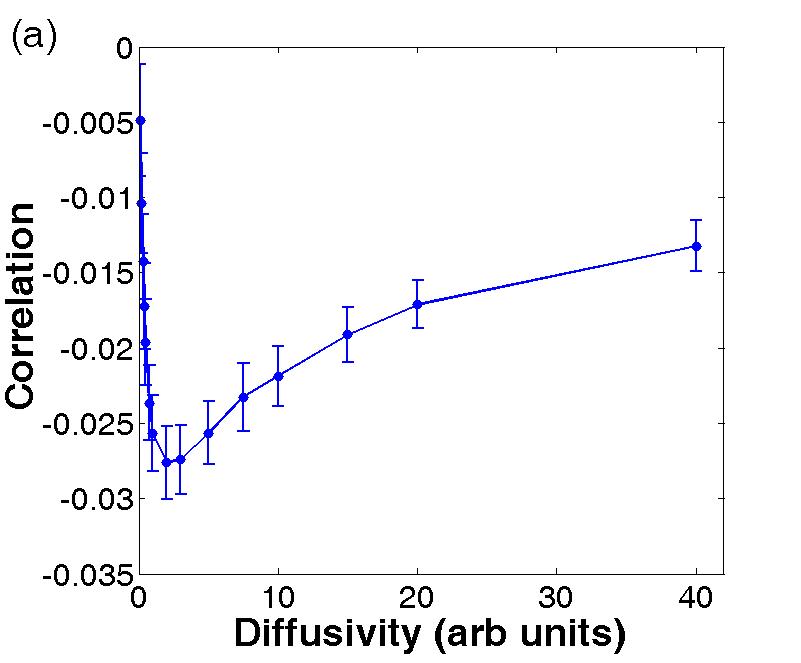}}
\centerline{\includegraphics[scale=0.26]{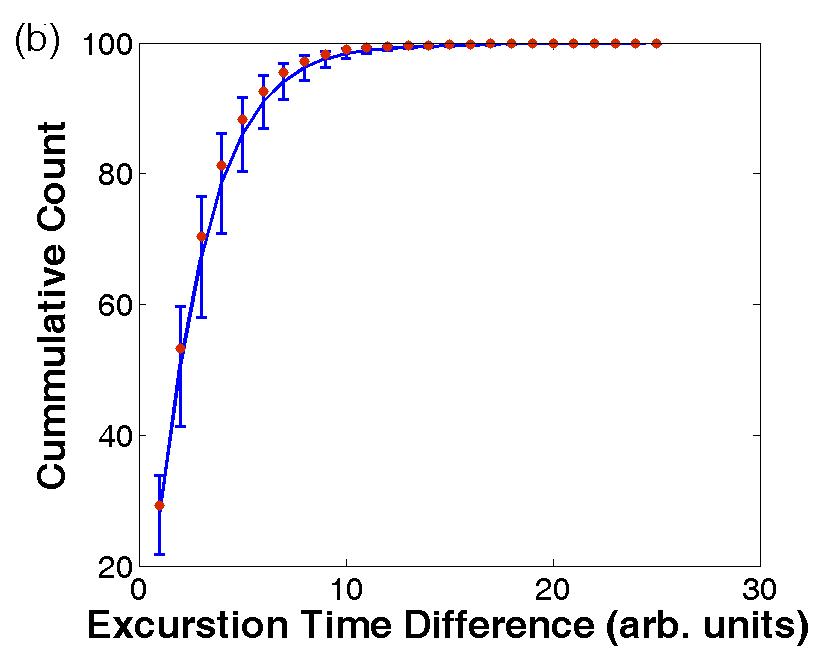}} \caption{ (a)
Correlation between ComK values of adjacent cells as a function of
diffusivity. The mutual inhibition, as identified by negative
correlation, vanishes for small diffusivities as well as for large
diffusivities. (b) Cumulative histograms of differences between
excursion times, of the entire colony (solid line) and adjacent
cells (markers).}\label{correlation}
\end{figure}

In the experimental data of \cite{elowitz}, sister cells showed no
correlation between their excursion times, once they become
competent. This led to the assumption that there was no important
spatial coupling. However, this conclusion might be inaccurate, as there might
be correlations between the cycles of neighboring cells but not in
their excursion times. In our simulations, excursion times only
slightly changed with the diffusion coefficient (about 10\% - data
not shown). And, no correlation was observed between the excursion
times of neighboring cells, as can be seen in Fig. 4(b), which is
very similar to the results of \cite{elowitz}. However, there are
notable correlations in the times of competence onset of neighboring
cells, as demonstrated in Fig. 4(a). This occurs because the $Q$
field has little effect on the competent state itself, but is
required to be high for the noise-induced excitation. Therefore, the
excursion times measurement is misleading and does not definitively indicate
whether there is, or there is not, a quorum sensing mechanism which
influences the competence transition. Clearly, our predictions could
be tested within the same experimental framework.

The existence of diffusing inhibition in this system suggests that
under certain circumstances the colony should be able to support a
patterned Turing state consisting of periodically repeating regions
of competent and non-competent cells. There are two distinct
possibilities; the system might have a Turing instability which
eliminates the vegetative state completely or it might be
multistable, meaning that patterned states co-exist with the uniform
one. We have found no evidence of instability, but multistability
can exist for large enough diffusion constant. In Fig. 5 we consider
the temporal evolution of a colony with an initial stripe pattern,
in which every tenth cell is competent. The values of the relevant
parameters are chosen to be: $a_k=0.004$, $b_k=0.14$, $k_0=0.2$,
$b_s=0.58$, $k_1=0.222$ and $c_s=0.1$. For these values the K-S
system of equations is bistable for $Q=1$ and vegetative monostable
for $Q<0.6$.  When the diffusion of Q is slow (Fig. 5(a)), the local
decrease in Q shifts the local K-S system of each competent cell to
the monostable, vegetative state, the cells revert to vegetative and
the pattern vanishes. But when diffusion is faster (Fig. 5(b)), the
averaged value of $Q$ can lie above the monostability threshold, and
the pattern is stable.
\begin{figure}
\hspace{.1cm} \centerline{\includegraphics[scale=0.3]{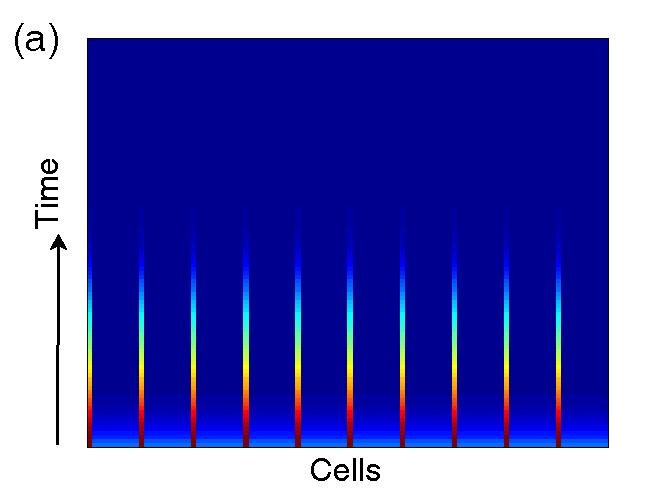}}
\centerline {\includegraphics[scale=0.3]{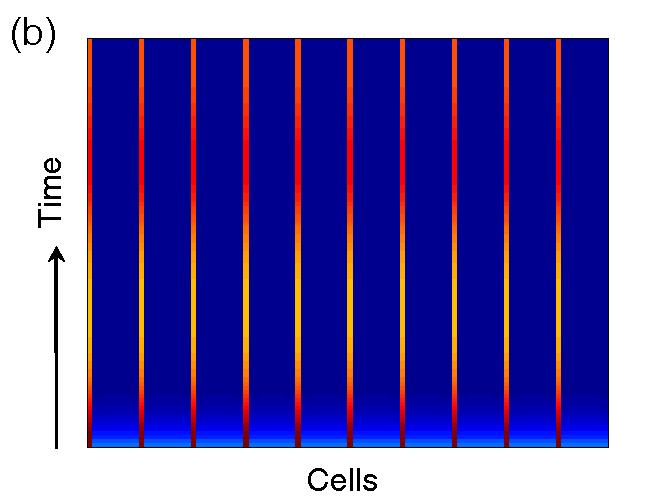}}
\vspace{-.75cm} \caption{A striped colony of cells Ð simulation
results.  The cell state is represented by color code (see  Fig. 3).
(a) For D=1, the stripes pattern is unstable and the cells return to
their vegetative state. (b) For D=2, the stripe pattern is stable.
}\label{stripe}
\end{figure}

We can get a better understanding of the stripe pattern and of the
lack of a Turing instability by explicitly evaluating the $Q$ field.
If we approximate the $\tanh$ in eq. 3 as a step function, we find
that in steady-state,
\begin{equation}
Q(x) = 1 - \frac{1}{2\sqrt{D \delta_Q}} \sum_i \exp{\left( -\sqrt{\frac{\delta_Q}{D}} | x -x_i| \right) }
\end{equation}
where the sum ranges over competent cells. This geometric series can
easily be calculated for any specific x; the result is shown in Fig.
6 as compared to the simulation data.
\begin{figure}
\centerline{\includegraphics[scale=0.4]{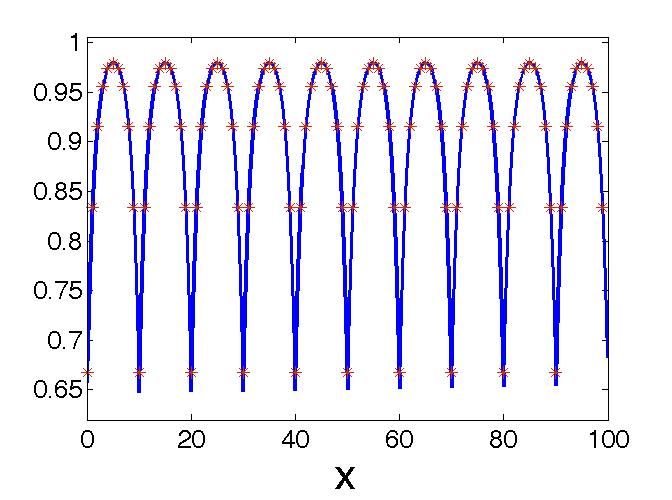}} \caption{The
steady state profile of Q(x), analytic solution(solid line)  and
simulation results (asterisks)}\label{Qfield}
\end{figure}
We see that the pattern is possible only because the self-inhibition is weakened sufficiently by the diffusion so as to allow the competent sites to remain in the bistable range. Paradoxically, the competent cells are the most inhibited, which is presumably why this pattern would not spontaneously appear from the uniform vegetative state. Studying this experimentally would necessitate preparing the system with an externally imposed pattern of CSF and then determining whether the pattern remains in place even after the external pattern is removed.

To summarize, we present a new model for collective behavior of a
colony of B. Subtilis cells. The condition of each cell is described
by the concentrations of two proteins, ComK and ComS. The dynamics
is described by a set of two rate equations, with a double-feedback
loop. The production rates effectively depend on the cell density in
the entire colony, sensed by a diffusible pheromone. This pheromone
is likely to depend on the protein ComK. This additional mechanism
creates an effective inhibition between neighboring cells, leading
to anti-correlation in the cells competent/vegetative cycles.
Excursion times of neighboring cells are found to be uncorrelated,
due to the self-excitatory mechanism of ComK, also in agreement with
experimental results. These findings can be experimentally verified
by the direct measurement of ComK concentration in the cells, as was
done in ref.~\cite{elowitz} via fluorescence techniques.

\begin{acknowledgments}
This work has been supported in part by the NSF-sponsored Center for
Theoretical Biological Physics (grant numbers PHY-0216576 and
PHY-0225630). EBJ was partially supported by the Tauber fund.
\end{acknowledgments}


\begin{thebibliography}{99}

\bibitem{review} D. Dubnau, {\em Ann. Rev. Microbiol.} {\textbf 53},
 217 (1999).
 \bibitem{cohen} E. Cohen, D. Kessler and H. Levine,
\prl \textbf{94}, 098102 (2005).

\bibitem{dubnau_bi} D. Dubnau and R. Losick, {\em Mol. Microbiol.} {\bf 61}, 564 (2006).
\bibitem{stochastic} R. Karmakar and I. Bose, arXiv:q-bio.QM/0702055v1(2007).
\bibitem{pos_feedback} H. Maamar and D. Dubnau, {\em Mol. Microbiol.} {\bf 56}, 615 (2005).
\bibitem{auto_stim} W.K. Smits,C.C. Eschevins, K.A. Susanna, S. Bron, O.P. Kuipers and L.W. Hamoen,
{\em Mol. Microbiol.} {\bf 56}, 604 (2005).
\bibitem{elowitz} G.M. Suel, J. Garcia-Ojalvo, L.M. Lieberman and M.B. Elowitz, {\em Nature} {\bf 440}, 545
(2006).
\bibitem{elowitz2} G. M. Suel, R. P. Kulkarni, J. Dworkin, J. Garcia-Ojalvo and M. B. Elowitz, {\em Science}, {\bf 315}, 1716 (2007).
\bibitem{excitable} A. Goldebeter, {\em Biochemical Oscillations and Cellular Rhythms},  Cambridge Univ. Press. (1996).
\bibitem{noise} M. B. Elowitz, A. J. Levine, E. D. Siggia and P. S. Swain, {\em Science} {\bf 202}, 1183 (2002).

\bibitem{paulsson} J. Paulsson, {\em Nature} {\bf 427}, 415 (2004).
\bibitem{alon} N. Rosenfeld, J. W. Young, U. Alon, P. S. Swain and M. B. Elowitz, {\em Science} {\bf 307}, 1962 (2005).
\bibitem{vano} J. M. Pedrazza and A. van Oudenaarden, {\em Science} {\bf 307}, 1965 (2005).

\bibitem{bassler} B. L. Bassler and R. Losick, {\em Cell} {\bf 125}, 237
(2006).
\bibitem{trends} E. Ben-jacob et al, {\it Trends in Microbiology} 12 (8) 366-372 (2004).
\bibitem{interface} E. Ben-Jacob and H. Levine, {\it J. Royal Soc. Interface} {\bf 3}, 197 (2006).
\bibitem{minsky} A. Minsky, private communication.

\bibitem{controlling_comp} L. W. Hamoen, G. Venema and O.P. Kuipers, {\em Microbiology} {\bf149}, 9 (2003).
\bibitem{signaling} B.A. Lazazzera,I.G. Kurster, R.S. McQuade and A.L. Grossman, {\em J. Bacteriology} {\bf 181}, 5193 (1999).
\bibitem{quorum} B. Lazezzera, {\em Curr. Opin. Microbiol.} {\bf 3}, 177 (2000).
\bibitem{schultz} D. Schultz, unpublished.
\bibitem{microarray} R.M. Berka, J. Hahn, M. Albano, I. Draskovic, M. Persuh, X. Ciu, A. Sloma, W. Widner and D. Dubnau, {\em Mol. Microbiol.} {\bf 43}, 1331 (2002).

\end{thebibliography}
\end{document}